\documentstyle[preprint,aps]{revtex}
\begin{document}
\draft
\tightenlines
\preprint{\vbox{\hbox{IFT--P.048/97}}}
%
\title{New Higgs Interactions in \boldmath{$ZZ\gamma$} and
\boldmath{$Z\gamma\gamma$} Production}
\author{S.\ M.\ Lietti and S.\ F.\ Novaes}
\address{Instituto de F\'{\i}sica Te\'orica, 
	 Universidade  Estadual Paulista, \\  
	 Rua Pamplona 145, CEP 01405--900 S\~ao Paulo, Brazil.}
%
\maketitle
\widetext
\begin{abstract}
The effect of new operators that only modify the bosonic
couplings of the Higgs boson, without altering the $WW\gamma$ or
$WWZ$ three--point functions, are examined in the $e^+ e^- \to
ZZ\gamma$ and $Z\gamma\gamma$ processes. We analyse the
constraints on these interactions that can be imposed by the
LEP II collider at CERN and at the Next Linear Collider.
\end{abstract}


\newpage 

\section{Introduction}

The experiments which are taking place at the CERN LEP II
electron--positron collider will be able to explore the gauge
structure of $WW\gamma$ and $WWZ$ three--point functions in the
$W$--pair production \cite{reviews}. Deviations from the
Standard Model (SM) predictions for this reaction would indicate
the existence of new physics effects. 

In general, such deviations can be parametrized in terms of
effective Lagrangians by adding to the SM Lagrangian,
high--dimension operators that describe the new phenomena
\cite{effective}. This model--independent approach accounts for
new physics that shows up at an energy scale $\Lambda$, larger
than the electroweak scale. The effective Lagrangians are
constructed with the light particle spectrum that exists at low
energies, while the heavy degrees of freedom are integrated out.
They are invariant under the $SU(2)_L \times U(1)_Y$ and, in the
linearly realized version, they involve, in addition to the usual
gauge--boson fields, also the light Higgs particle. The most
general dimension--6 effective Lagrangian, containing all SM
bosonic fields, that is  $C$ and $P$ even, was constructed in
Ref.\ \cite{hisz}.

In this letter, we explore the consequence of new operators that
give rise to anomalous Higgs boson couplings, without affecting
the self--interaction amongst the gauge bosons. These anomalous
interactions can {\it not} be constraint by the LEP II results on
W--pair production. In particular, we study the reactions $e^+
e^- \to ZZ\gamma$ and $Z\gamma\gamma$. These processes do not
involve any triple--vector--boson coupling, but only $HZ\gamma$
and $H\gamma\gamma$ anomalous Higgs interactions. In the SM,
neglecting the Higgs--electron coupling, there are no Higgs
contributions to these reactions at tree level since the
$HZ\gamma$ and $H\gamma\gamma$ couplings are generated only at
one--loop \cite{hgg,hgz}. 

Out of the eleven independent operators constructed in Ref.\
\cite{hisz}, two of them describe new interactions between the
Higgs particle and the vector bosons, 
\begin{equation}
{\cal L}_{\text{eff}} =  \frac{f_{WW}}{\Lambda^2} \;
\Phi^{\dagger} \hat{W}_{\mu \nu} \hat{W}^{\mu \nu} \Phi + 
\frac{f_{BB}}{\Lambda^2} \; \Phi^{\dagger} \hat{B}_{\mu \nu}
\hat{B}^{\mu \nu} \Phi  \; ,
\label{lagrangian}
\end{equation}
where, in the unitary gauge, the Higgs doublet becomes $\Phi =
(1/\sqrt{2}) [\, 0 \, , \, (v + H) \, ]^T$, and  $\hat{B}_{\mu
\nu} = i (g'/2) B_{\mu \nu}$, and  $\hat{W}_{\mu \nu} = i (g/2)
\sigma^a W^a_{\mu \nu}$, with $B_{\mu \nu}$ and $ W^a_{\mu \nu}$
being the field strength tensors of the $U(1)$ and $SU(2)$ gauge
fields respectively. 

We should notice that the operators (\ref{lagrangian}) contribute
only to the anomalous Higgs couplings, $HVV$, $V = W, Z, \gamma$,
since their possible contribution to the self--gauge--boson
couplings, $WW\gamma$ and $WWZ$, can be completely absorbed in
the redefinition of the SM fields and gauge couplings
\cite{hisz}. Furthermore, no new $ZZ\gamma$ or $Z\gamma\gamma$
vertices, like the ones studied in Ref.\ \cite{baur}, are
generated via these operators.   Therefore, studies of anomalous
trilinear gauge boson couplings are insensitive to the
coefficients $f_{WW}$ and $f_{BB}$. Anomalous Higgs boson
couplings have already  been studied in Higgs and Z boson decays
\cite{hagiwara2}, in $e^+ e^-$ \cite{ee,our}, $\gamma\gamma$
\cite{gamma}, and $p\bar{p}$ colliders \cite{fer}. Here, in order
to impose limits on the dimension--6 operators (\ref{lagrangian})
that generate the new Higgs interactions, we examine the total
$ZZ\gamma$ and $Z\gamma\gamma$ yield and kinematical
distributions of the final state particles at LEP II and at the
Next Linear Collider, NLC.

\section{Results for the Anomalous $ZZ\gamma$ and $Z\gamma\gamma$
Production}

In our analysis of the $e^+ e^- \to  Z Z \gamma$ and $Z \gamma
\gamma$ reactions, we have taken into account the standard one
loop Higgs contributions \cite{hgg,hgz} to these processes. In
this way, the $Z\gamma\gamma$ production involves seven (eight)
standard (anomalous) Feynman contributions, while in the
$ZZ\gamma$ process there are ten (twelve) standard (anomalous)
diagrams. In order to compute these contributions, we have
incorporated  all anomalous couplings in Helas--type \cite{helas}
Fortran subroutines. These new subroutines were used to adapt a
Madgraph \cite{madgraph} output to include all the anomalous
contributions. We have checked that our code passed the
non--trivial test of electromagnetic gauge invariance.

We investigate the $Z\gamma\gamma$ and  $ZZ\gamma$ production
both at LEP II and at NLC, assuming a center--of--mass  energy of
$\sqrt{s} = 190$ and 500 GeV and an integrated luminosity of
${\cal L} = 0.5$ and 50 fb$^{-1}$ respectively.  We required that
the photon energy, $E_\gamma$, is larger than 5 GeV for LEP II
and than 20 GeV for NLC. We applied an isolation cut by requiring
that the angle between any two final state particles is larger
than $15^\circ$. The same cut was applied to the angle of the
bosons with the beam pipe. 

Our results for the sensitivity of LEP II to the operators
appearing in the effective Lagrangian (\ref{lagrangian}) indicate
that the reaction $e^+ e^- \to Z \gamma \gamma$ can provide
limits on both anomalous couplings since
$\sigma_{Z\gamma\gamma}^{SM} = 654$ fb, and we can expect around
300 events per year.  A 1$\sigma$ deviation in the total cross
section, assuming a Higgs boson with mass of 100 GeV, requires
that,
\begin{equation}
-296 < \frac{f_{BB}}{\Lambda^2} < 500 \;\; \text{TeV}^{-2} \;\;,\;\;  
-58 < \frac{f_{WW}}{\Lambda^2} < 92 \;\; \text{TeV}^{-2} \; . 
\label{bound:lep}
\end{equation} 
These results are similar to those arising from the analysis
of the reaction $e^+ e^- \to W^+ W^- \gamma$ \cite{our2}, where a
1$\sigma$ deviation implies $|f_{WW}| \lesssim 150$. 

On the other hand, the reaction $e^+ e^- \to ZZ\gamma$ is almost
insensitive to the anomalous coefficient $f_{BB}$.  For a Higgs
boson mass of $100$ GeV, a 1$\sigma$ deviation can be observed
for $|f_{WW}/\Lambda^2| \sim 300$ TeV$^{-2}$.  The small SM cross
section for this process, $\sigma_{ZZ\gamma}^{SM} = 4.4$ fb,
leads to very few events per year at LEP II, and in consequence,
the constraints from this reaction are limited by the poor
statistics.

Due to the increase on the available phase space, we are able to
establish tighter constraints on the coefficients $f_{WW,BB}$ by
studying the contribution of the operators (\ref{lagrangian}) to
these same processes at NLC. For instance, at $\sqrt{s} = 500$
GeV, $\sigma_{Z \gamma \gamma}^{SM} =  58$ fb, which yields 2900
events per year assuming the expected luminosity for NLC (${\cal
L} = 50$ fb$^{-1}$).  Requiring a maximum deviation of 2$\sigma$
in the total cross section, and assuming a Higgs boson of 200
GeV, we obtain the allowed  ranges $-39 < f_{BB}/\Lambda^2 < 35$
TeV$^{-2}$ and $-9.6 < f_{WW}/\Lambda^2 < 14$ TeV$^{-2}$.

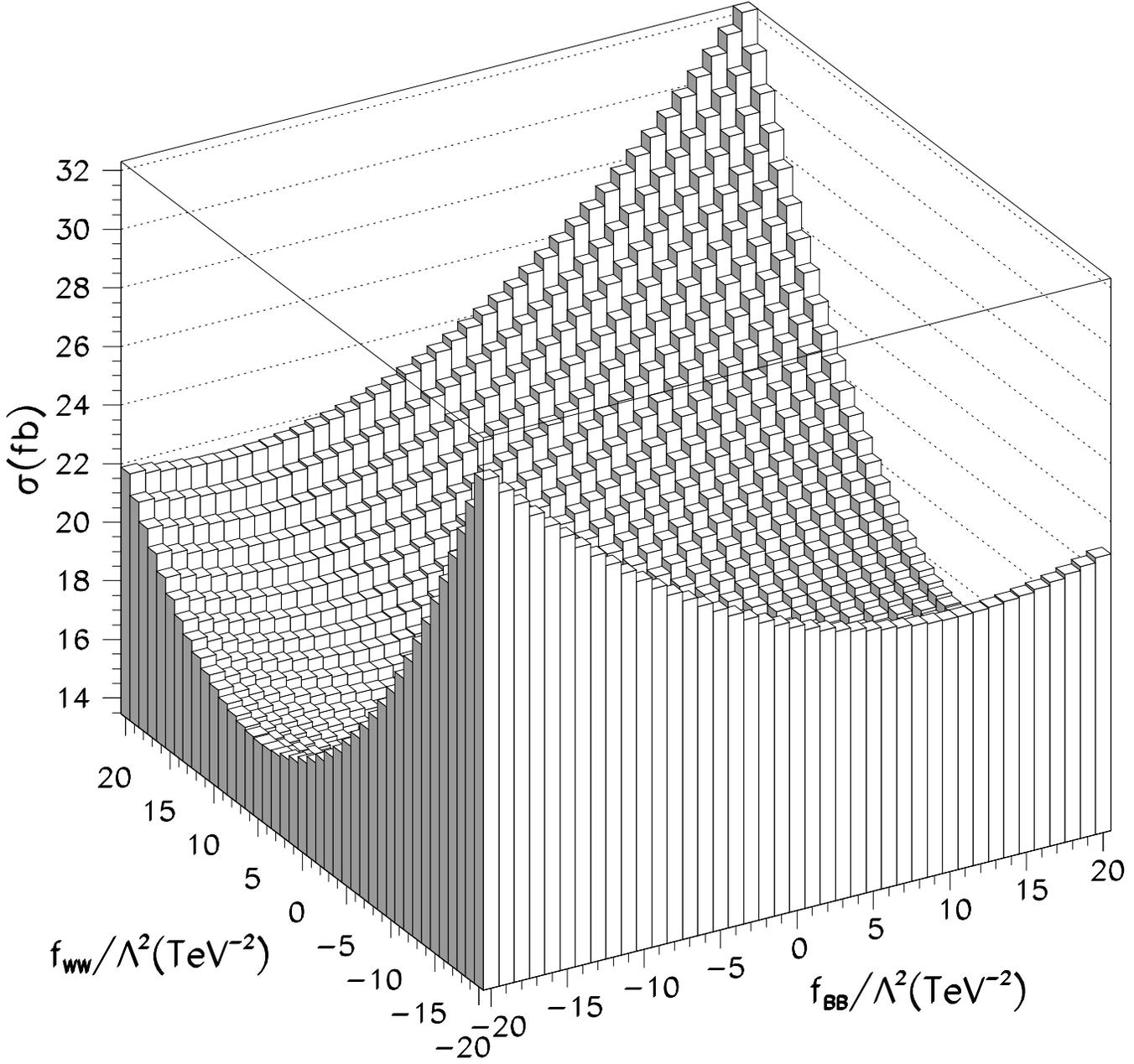
\begin{figure}
\protect
\caption{Total cross section for the reaction $e^+ e^- \to Z Z
\gamma$ at NLC, for $M_H =200$ GeV, as a function of $f_{BB}$ and
$f_{WW}$.}
\label{fig:1}
\end{figure}

Unlike at LEP II, at NLC the best constraint on the anomalous
couplings come from the reaction $e^+ e^- \to Z Z \gamma$, which
gives rise to $\sim 675$ events per year at $\sqrt{s} = 500$ GeV.
Figure \ref{fig:1} shows the effect  of the anomalous operators
(\ref{lagrangian}) in the total cross section of this reaction. A
2$\sigma$ deviation in the total cross section sets the bounds,  
\begin{equation}
-12.6 < \frac{f_{BB}}{\Lambda^2} < 8.5 \;\; \text{TeV}^{-2} \;\;,\;\;  
-6.6 <  \frac{f_{WW}}{\Lambda^2} < 5.7 \;\; \text{TeV}^{-2} \; ,
\label{bound:gzznlc}
\end{equation}
also for a 200 GeV Higgs boson. Furthermore, we can expect the
new interactions to affect most the longitudinally polarized
gauge bosons production due to the presence of the scalar Higgs
boson as intermediate state.  We tried to take advantage of this
fact by studying the longitudinal $Z$--pair production which
brings on a slightly better result, {\it i.e.\/} $-11.9 <
f_{BB}/\Lambda^2 < 7.8$ TeV$^{-2}$ and $-6.4 < f_{WW}/\Lambda^2 <
5.2$ TeV$^{-2}$ for a 2$\sigma$ effect. Unfortunately the
improvement is very small because the requirement of polarized
$Z$ reduces by two orders of magnitude the total yield.

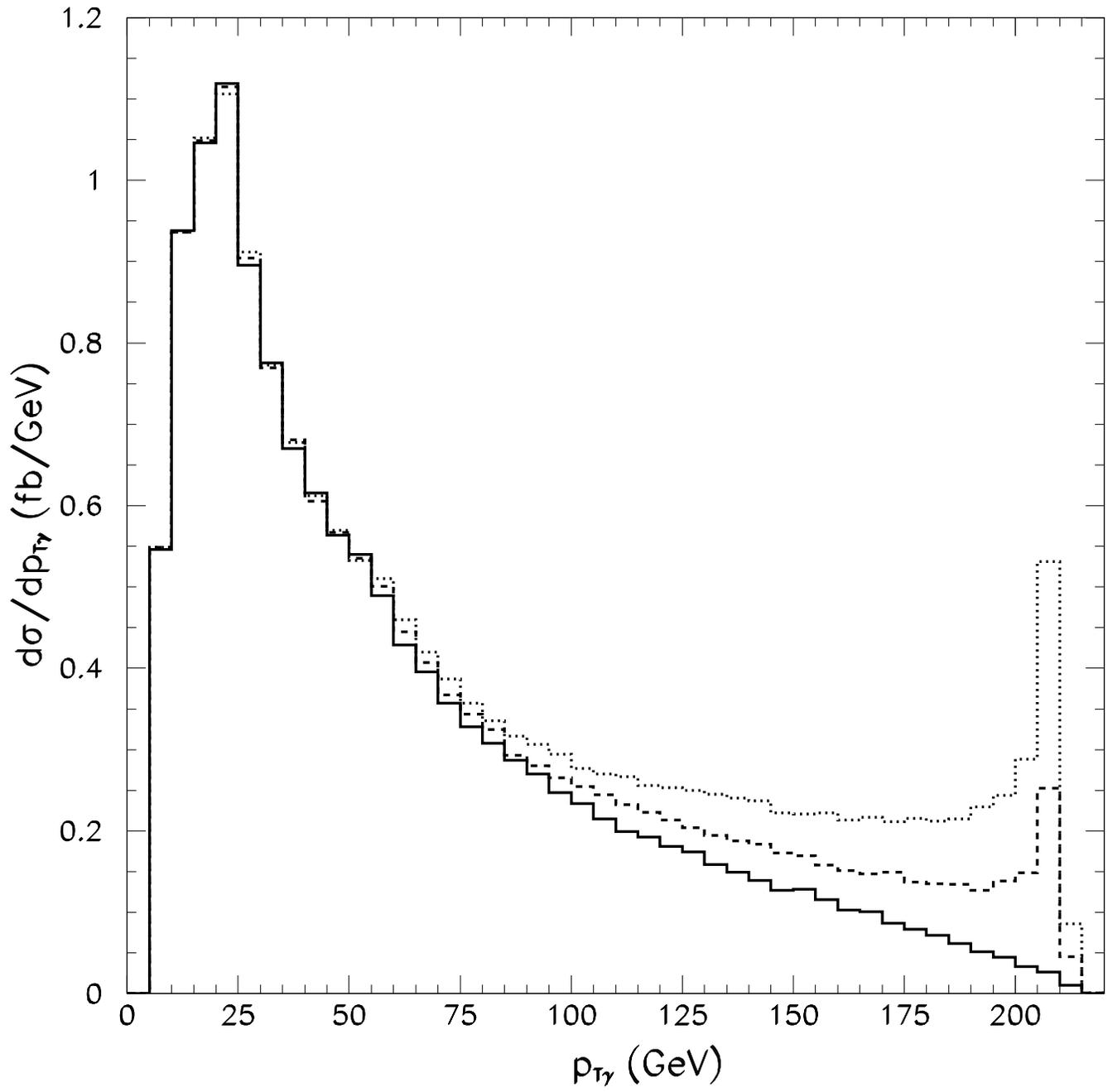
\begin{figure}
\protect
\caption{Photon transverse momentum distribution for  $e^+ e^- \to
Z Z \gamma$ at NLC, assuming $M_H = 200$ GeV, for the Standard
Model (solid line), and for  $f_{BB}/\Lambda^2 = 10$  and
$f_{WW}/\Lambda^2 =0$ TeV$^{-2}$ (dashed line) and for
$f_{WW}/\Lambda^2 = 10$ and $f_{BB}/\Lambda^2 =0$ TeV$^{-2}$
(dotted line).}
\label{fig:2}
\end{figure}

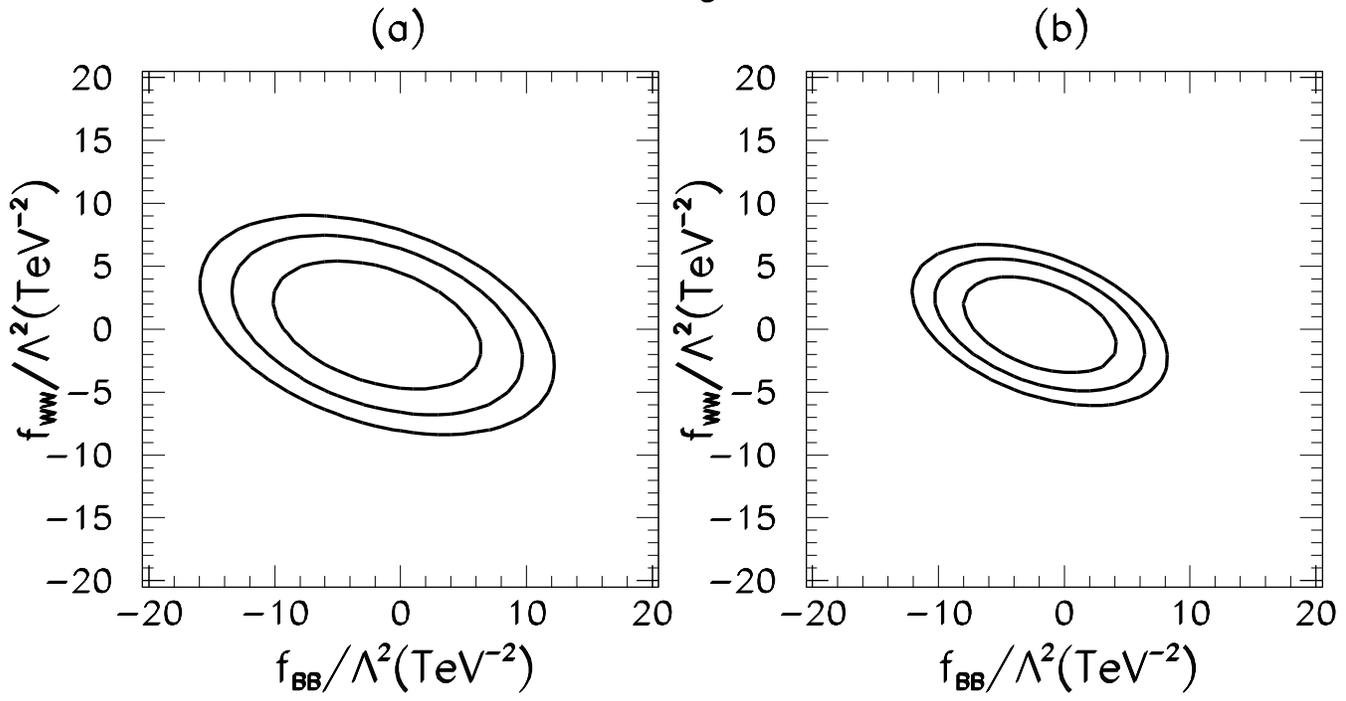
\begin{figure}
\protect
\caption{Contour plot of $f_{BB} \times f_{WW}$, from  $e^+ e^-
\to Z Z \gamma$ at NLC, for $M_H =200$ GeV. The curves show  the
one, two, and three sigma deviations from the Standard Model
total cross section: (a) No cut on $p_{T_\gamma}$, and (b) with a
cut of $p_{T_\gamma} > 100$ GeV.}
\label{fig:3}
\end{figure}

We have also investigated different distributions of the final
state particles in order to search for kinematical cuts that could
improve the NLC sensitivity. The most promising variable is the
photon transverse momentum which distribution is presented in
Fig.\ \ref{fig:2}  for the unpolarized case. We observe that the
contribution of the anomalous couplings is larger in the high
$p_{T_\gamma}$ region. Therefore a cut of $p_{T_\gamma} > 100$
GeV drastically reduces the background. The improvement on
$f_{WW,BB}$ bounds can be clearly seen from Fig.\ \ref{fig:3}
where a 1, 2, and 3$\sigma$ deviations in the total cross section
is shown before and after the above cut is applied.

The contribution of the anomalous couplings is dominated by
on--mass--shell Higgs production with the subsequent $H \to Z Z$
decay. The peak around $210$ GeV in the photon transverse
momentum distribution is due to the monochromatic photon with
$E_\gamma^{\text{mono}}= (s - M_H^2)/(2 \sqrt{s})$ which is
emitted in association with the 200 GeV Higgs boson. Therefore
the best constraints are obtained at NLC for Higgs boson masses
in the range $2 M_Z \leq M_H \leq  (\sqrt{s} -
E_\gamma^{\text{min}})$ GeV, where a on--shell production is
allowed. 

We present in Table \ref{tab:1} the limits on the coefficients
$f_{BB}$ and  $f_{WW}$ based on a $2\sigma$ deviation in the
total cross section for a Higgs mass in the range $200  \leq M_H
\leq 350$ GeV. Table \ref{tab:2} shows the same limits after the
cut in the photon transverse momentum distribution of
$p_{T_\gamma} > 100$ GeV is implemented. These results are
comparable to the limits for $f_{BB}$ and $f_{WW}$ obtained from
the study of  the reaction $e^+ e^- \to W^+ W^- \gamma$ presented
in  Ref.\  \cite{our2}, suggesting that the processes considered
here can be used as a complementary tool in the study of
anomalous Higgs interactions.

In conclusion, the search for the effect of higher dimensional
operators that give rise to anomalous Higgs boson couplings may
provide important information on physics beyond the Standard
Model and should be pursued in all possible reactions. We have
studied here the triple vector boson production, $ZZ\gamma$ and
$Z\gamma\gamma$, in $e^+ e^-$ colliders focusing on the operators
that generate only anomalous Higgs boson couplings and cannot be
tested in the $W$--pair production.  We established the limits
that can be imposed at LEP II and NLC through the analysis of the
deviations on the total cross section.  Furthermore, we used the
photon transverse momentum spectrum and polarization of the $Z$'s
to improve these constraints to few TeV$^{-2}$ at the NLC.

\acknowledgments
We would like to thank M.\ C.\ Gonzalez--Garcia for very useful
discussions. This work was supported by Conselho Nacional de
Desenvolvimento Cient\'{\i}fico e Tecnol\'ogico (CNPq), and by
Funda\c{c}\~ao de Amparo \`a Pesquisa do Estado de S\~ao Paulo
(FAPESP).


\begin{table}
\begin{tabular}{||c||c||c||}
$M_H$(GeV) & $f_{BB}/\Lambda^2$ & $f_{WW}/\Lambda^2$ \\
\hline 
\hline
200 & ( $-$12.6 , 8.5 ) & ( $-$6.6 , 5.7 ) \\
\hline
250 & ( $-$13.3 , 9.5 ) & ( $-$7.5 , 6.2 ) \\ 
\hline
300 & ( $-$16.0 , 12.5 ) & ( $-$9.1 , 7.9 ) \\ 
\hline
350 & ( $-$21.9 , 18.3 ) & ( $-$12.3 , 11.5 )    
\end{tabular}
\caption{Range of the allowed values of the  coefficients
$f_{BB}$ and $f_{WW}$, in  TeV$^{-2}$, for a $2\sigma$ deviation
in the total cross section of the process $e^+ e^- \to Z Z
\gamma$ at NLC.}
\label{tab:1}
\end{table}
\begin{table}
\begin{tabular}{||c||c||c||}
$M_H$(GeV) & $f_{BB}/\Lambda^2$ & $f_{WW}/\Lambda^2$ \\
\hline 
\hline
200 & ( $-$9.6 , 5.4 ) & ( $-$5.4 , 3.7 ) \\
\hline
250 & ( $-$10.3 , 6.5 ) & ( $-$5.9 , 4.4 ) \\ 
\hline
300 & ( $-$12.9 , 9.3 ) & ( $-$7.3 , 5.9 ) \\ 
\hline
350 & ( $-$20.5 , 16.1 ) & ( $-$10.6 , 10.1 )    
\end{tabular}
\caption{The same as Table \protect\ref{tab:1} after the
implementation of the cut $p_{T_\gamma} > 100$ GeV.}
\label{tab:2}
\end{table}

\end{document}